\documentstyle[12pt,aasms,flushrt]{article}

\def\be{\begin{equation}}
\def\ee{\end{equation}}
\def\br{{\bf r}}

\begin{document}

\title{\bf The Origin of Chaos in the Outer Solar System}

\medskip

\author{N. Murray$^1$ and M. Holman$^2$}

\medskip

\affil{$^1$Canadian Institute for Theoretical Astrophysics, \\
60 St. George st.,  University of Toronto, Toronto, Ontario M5S 3H8, Canada}

\medskip

\affil{$^2$Harvard-Smithsonian Center for Astrophysics,\\
60 Garden st., Cambridge, MA 02138, U.S.A.}

\medskip

\vfil\eject
%\begin{abstract}
Classical analytic theories of the solar system indicate that it is
stable, but numerical integrations suggest that it is chaotic. This
disagreement is resolved by a new analytic theory. The theory shows
that the chaos among the Jovian planets results from the overlap of
the components of a mean motion resonance among Jupiter, Saturn, and
Uranus, and provides rough estimates of the Lyapunov time (10
million years) and the dynamical lifetime of Uranus ($10^{18}$ years). The
Jovian planets must have entered the resonance after all the gas and
most of the planetesimals in the protoplanetary disk were removed.
%\end{abstract}

\vfil\eject

\noindent
The predictability of planetary motions was
largely responsible for the acceptance of Newton's theory of
gravitation. In spite of this, Newton doubted the long term stability
of the solar system. Laplace noted that the ratios $\mu$ of planetary
masses $M$ to solar mass $M_\odot$ are small ($\mu\sim10^{-3}$ to
$10^{-9}$), as are the planet's orbital eccentricities $e\sim10^{-2}$
and inclinations $i\sim10^{-2}$ (in radians).  Neglecting terms
proportional to second or higher powers of these quantities, Laplace
showed that the motions of the planets were stable \cite{stable}.  In
this century, Arnold showed that for $\mu$, $e$, and $i$ of
order $10^{-43}$, most planetary systems, in the sense of measure
theory, are stable and undergo quasiperiodic but bounded variations in
semimajor axis ($a$), $e$, and $i$ for each planet\cite{Arnold}. 
However, the values of the small quantities in our
solar system are much larger than $10^{-43}$, so the
applicability of Arnold's theory is uncertain.  Studies over the last
decade have been dominated by brute force numerical
integration. Sussman and Wisdom \cite{SW}, and Laskar \cite{Laskar},
performed numerical integrations of the planet's orbits and found
positive Lyapunov exponents, indicating
that they are chaotic. Sussman and Wisdom also showed that
integrations of the Jovian planets Jupiter, Saturn, Uranus, and
Neptune are by themselves chaotic. In neither case are the variations
in $a$, $e$ and $i$ quasiperiodic, nor is it clear that they are
bounded.  Are the numerical results incorrect, or are the classical
calculations simply inapplicable?

We show analytically that the results of Laplace and Arnold do not
apply to our solar system. The chaos seen in integrations of the outer
planets arises from the overlap of the components of a three body mean
motion resonance among Jupiter, Saturn, and Uranus, with a minor
role played by a similar resonance among Saturn, Uranus, and
Neptune.  We test the theory using a suite of numerical
integrations. The widths $\Delta a/a$ of the individual resonances are
of order $3\times 10^{-6}$, so that small changes in the initial
conditions of the planets can lead to regular motion. This explains 
the puzzling dependence of Lyapunov time with integration step size
seen in the outer planet integrations of  Sussman and Wisdom
\cite{SW}. 
However, the uncertainties in
the initial conditions, and those introduced by our numerical model,
are comfortably smaller than the width of the individual resonances,
so our solar system is almost certainly chaotic. The resonance is
extremely weak and hence easily disrupted. Torques exerted on the
planets by the protoplanetary gas disk and by planetesimals were
orders of magnitude larger than the resonant torques, so most of the
planetesimals and all the gas must have been cleared from the outer
solar system before the planets entered the resonance.

{\bf Analytic Theory.} {\it Orbital dynamics.}
Planets in the solar system follow nearly Keplerian
orbits$^*$\footnotetext{The orbits have sizes and shapes described by
semimajor axis $a$ and eccentricity $e$. The orientation of an orbit
is described by the inclination $i$, the longitude of the ascending
node $\Omega$, and the longitude of perihelion $\varpi$, while the
location of the planet in the orbit is described by the mean anomaly
$l$ or equivalently the mean longitude $\lambda\equiv
l+\varpi$. Collectively these variables are called orbital
elements. In the Kepler problem, where a single planet orbits a
spherical star, all the elements of the planet except the mean
longitude are fixed, which is why the elements are useful
quantities. We use the masses and $a$, $e$, and $i$ from the JPL
ephemeris DE200 (Table 1)}.  The orbit of each planet can be thought
of as consisting of three nonlinear oscillators, 
corresponding to the three spatial directions. The Kepler problem is
unusual in that all three oscillations have the same frequency. The
orbital elements were chosen to take advantage of this degeneracy. The
angle $l$ varies on the orbital time scale, while the angle $\varpi$
describing radial motion and the angle $\Omega$ describing vertical
motion are fixed. In the actual solar system $\varpi$ and $\Omega$ are
time dependent, with frequencies denoted by $g_j$ 
and $s_j$ respectively. These frequencies are proportional to
the mass ratios $\mu$, and are consequently much smaller than the mean
motion $n=dl/dt$, the time rate of change of the mean anomaly. While
our model contains only the Jovian planets, we label  $g_j$ and $s_j$
with $j=5$, $6$, $7$, and $8$, corresponding to the radial order of
the planets in the solar system. The mean motions $n$ in units of
cycles per day and the modal frequencies of the Jovian planets were
determined by numerical integration of the equations of motion
(Table 2). Each planet's elements vary with all the frequencies $s$ and
$g$. For example
\be \label{secular}%$
e_J\sin\varpi_J\approx e_{55}\sin(g_5t+\xi_5)+e_{56}\sin(g_6t+\xi_6)+\ldots,
\ee %$
where $e_{55}\approx0.044$, $e_{56}\approx 0.016$, and $\xi_5$ and
$\xi_6$ are constants.

{\it Resonances and chaos.} A resonance occurs when two or more
oscillators are coupled in such a 
way that a linear combination of their angles
$\sigma\equiv\sum_ip_i\theta_i$ undergoes a bounded oscillation, in
which case $\sigma$ is said to librate. In the sum defining $\sigma$,
$i$ denotes the $i$th oscillator and the $p_i$s are (possibly
negative) integers. When the oscillators are not resonant, all
possible combinations of $\theta_i$'s increase or decrease
indefinitely, in which case $\sigma$ is said to rotate.  The physical
significance of a resonance is that energy is exchanged between the
oscillators over a libration period, which is large compared to the
oscillation period of any of the oscillators. This prolonged exchange
can lead to large changes in the motion of the system. The orbit that
divides regions of phase space where $\sigma$ librates from those
where $\sigma$ rotates is called the separatrix.

The other bit of dynamics needed to understand our result is the notion
of resonance overlap. Chaos in Hamiltonian systems, of which the
motions of the planets are an example, arises when the separatrix of
one resonance is perturbed by another resonance. The extent of the
chaos depends on the stochasticity parameter $K$, which is a function
of the of separatrix width divided by the distance between resonances. If $K$
is small, there is little chaos, but for $K>1$ the region in the
immediate vicinity of the resonances is primarily chaotic
\cite{Chirikov}. An orbit which, at different times, both librates and
rotates must cross the separatrix, and is therefore chaotic. Another
signature of chaos is that two initially nearby chaotic orbits diverge
exponentially with time; in our numerical work we use both diagnostics.

{\it Two body mean motion resonances.} Two planets are said to be in a
mean motion resonance when $p_1d\lambda_1/dt\approx p_2d\lambda_2/dt$. In
that case conjunctions between the planets occur at nearly fixed
locations in space. The designation ``mean motion'' is a little
misleading, because if $p_1\ne p_2$ there is no coupling between the $(\lambda,a)$
motion of two planets that does not involve a third degree of freedom,
either the radial $(\varpi,e)$ or vertical $(\Omega,i)$ motion of at
least one of the planets\cite{HM}. 

There are no two body mean motion resonances among the planets.
However, there is a near mean motion resonance between Jupiter and
Saturn; Jupiter makes five circuits around the sun in about the same
time that Saturn orbits twice. Saturn affects the orbit of Jupiter through
its gravity, described by the potential
\be %$
\phi=-{GM_S\over|\br_J-\br_S|}, 
\ee %$ 
where $M_S$ is the mass of Saturn, $\br_J$ and $\br_S$ are the position
vectors of Jupiter and Saturn, and $G$ is the gravitational constant.
To see the resonance mathematically, we expand $\br_J$ and $\br_S$ in terms
of the orbital elements of the two planets, keeping only the lowest
order terms:
\be \label{great_inequality}%$
\phi=-{GM_S\over a_S}
\sum_{k,q,p,r}\phi_{k,q,p,r}^{(2,5)}(a_S/a_J)e_S^ke_J^qi_S^pi_J^r\cos
\left[2\lambda_J-5\lambda_S+k\varpi_S+q\varpi_J+p\Omega_S+r\Omega_J\right].
\ee %$
The amplitudes $\phi_{k,p,q,r}$ can be found in classic references
\cite{Leverrier}.  This result shows explicitly that the gravitational
coupling between two bodies on Keplerian orbits always involves either
$(\varpi,e)$ or $(\Omega,i)$, so that at least three oscillators are
affected.  Symmetry considerations show that the integers in the
argument of the cosine must sum to zero, $2-5+k+q+p+r=0$, and that
$p+r$ must be even.  To lowest order in the eccentricities and
inclinations, the integers $k$, $q$, $p$, and $r$ are non-negative and
must sum to three. The strength of the coupling is proportional to
$e^3$ or $ei^2$, so this resonance is said to be of third order. Hence
there are ten frequencies associated with the resonance, four
involving only perihelion precession rates, such as
\be %$
2\dot\lambda_J-5\dot\lambda_S+2\dot\varpi_J+\dot\varpi_S,
\ee %$
and six involving the precession rates of the nodal lines, including
\be %$
2\dot\lambda_J-5\dot\lambda_S+\dot\varpi_J+\dot\Omega_J+\dot\Omega_S.
\ee %$
The dot over the angles in these expressions denotes a time derivative.
Each of the ten members of Eqn. (\ref{great_inequality}) is
referred to as a resonant term or, sometimes, as a resonance.  The reason
for this misuse of terminology is that, while none of the
frequencies associated with these terms in our solar system vanish,
they are much smaller than the mean motions of Jupiter and Saturn. As
a result, the resonant terms have a strong effect on the orbits of the
two planets. 

Eighteenth century astronomers, unaware of the significance of these
long period terms, noted a discrepancy between the predicted and
observed longitude of Jupiter and Saturn. This discrepancy, known as
the great inequality \cite{Moulton}, was finally explained by
Laplace. He noted that the resonant terms given by
Eqn. (\ref{great_inequality}) force a periodic displacement of $21$
minutes of arc in Jupiter's longitude and 49 minutes of arc in
Saturn's, showing that the predictions of the law of gravitation
agreed with observations of the two planets.

The largest effect of Saturn's gravity on $e_J$ is the secular
variations described by Eqn. (\ref{secular}). However, the most
relevant component of Saturn's gravity for chaotic motion is described
by Eqn. (\ref{great_inequality}). This component forces much smaller
variations in $e_J\sin\varpi_J$ given by
\begin{eqnarray} \label{ej} %$
e_J^{2,5}\sin\varpi_J&\approx& {\mu_S\over(2-5n_S/n_J)}{a_J\over a_S}\sum_{p>0} 
\phi_{k,p,q,r}^{(2,5)} e_S^k e_J^{p-1}i_J^q i_S^r\nonumber \\
& & \times\sin[2\lambda_J-5\lambda_S+k\varpi_S+(p-1)\varpi_J+q\Omega_J+r\Omega_S].
\end{eqnarray} %$
The largest variation in $e_J^{2,5}$, corresponding to $k=2$,
$p-1=q=r=0$ and $\phi_{2,1,0,0}\approx 9.6$, has an amplitude of about
$3.5\times10^{-4}$. Our numerical integrations yield
$3.7\times10^{-4}$, consistent within the errors introduced by keeping
only the highest order term in $e$. This variation in $e_J$ plays a
central role in producing chaos among the outer planets.

There are other two body near mean motion resonances in the solar
system. Of particular 
relevance here is the $7\lambda_U-\lambda_J$ near resonance between
Jupiter and Uranus. The potential experienced by Uranus is
\be \label{seven2one}%$
\phi=-{GM_J\over
a_U}\sum_{k,q,p,r}\phi_{k,q,p,r}^{(7,1)}e_J^ke_U^qi_J^pi_U^r\cos
\left[\lambda_J-7\lambda_U+k\varpi_J+q\varpi_U+p\Omega_J+r\Omega_U\right].
\ee %$
To lowest order (sixth) in $e$ and $i$, there are 44 terms. The
coefficients $\phi_{k,q,p,r}^{(7,1)}$ range from  $\sim10^{-3}$ to
$\sim10$. By itself this resonance has little effect on the dynamics of
the solar system. 

{\it Three body mean motion resonances.} Now consider the fact that
$e_J\sin\varpi_J$ varies; 
substituting (\ref{ej}) into (\ref{seven2one}), we find the potential
experienced by Uranus due to the non-Keplerian orbit of Jupiter;
\begin{eqnarray}\label{potential357} %$
\phi&\approx&-{GM_J\over a_U}\mu_S\epsilon^{-1}_{JS}\alpha_{JS}\sum_{p=0}^5(6-p)
\phi^{(7,1)}_{6-p,p,0,0}\phi^{(5,2)}_{2,1,0,0} e_J^{5-p} e_U^p e_S^2\nonumber\\
&&\times\sin[3\lambda_J-5\lambda_S-7\lambda_U+7\varpi_J+p(\varpi_U-\varpi_J)+2\varpi_S],
\end{eqnarray} %$
where $\alpha_{JS}=a_J/a_S\approx 0.55$ and
$\epsilon_{JS}=|2-5(n_S/n_J)|\approx1.3\times10^{-2}$. For simplicity
we have ignored terms involving the inclinations and kept only terms
proportional to $e_S^2$. This three body mean motion resonance is
second order in the masses of the planets (both $\mu_J$ and $\mu_S$
appear) and seventh order in $e$.

Using the frequencies in Table 1, and accounting for terms involving
$i$, we find a mixed $e$-$i$ resonance at $a_U\approx19.21796\hbox{
AU}$ associated with the argument
\be %$
3\lambda_J-5\lambda_S-7\lambda_U+7g_6t+2s_7t
\ee %$
We find a cluster of eccentricity resonances centered at
$a_U\approx19.2163\hbox{ AU}$  with argument
\be %$
3\lambda_J-5\lambda_S-7\lambda_U+(2-q)g_5t+7g_6t+qg_7t,
\ee %$
where $0\le q\le 2$, and at $a_U\approx19.2193\hbox{ AU}$ with argument
\be \label{resonance}%$
3\lambda_J-5\lambda_S-7\lambda_U+(3-q)g_5t+6g_6t+qg_7t,
\ee %$
where $0\le q\le 3$. At the present epoch the JPL ephemeris DE200 has
$a_U\approx19.21895\hbox{ AU}$ 

For simplicity we have described only one type of term in the potential
experienced by Uranus, that due to the influence of Jupiter as it
moves in the potential of the Sun and Saturn, as reflected in
$e_J\sin\varpi_J$. There are similar contributions to the potential due to
variations in Jupiter's other orbital elements. Furthermore,
there are weaker resonances due to the gravity of Saturn, moving on an orbit 
perturbed by Jupiter, and acting on Uranus. Finally there are much
smaller terms due to the direct perturbations of Uranus by Saturn and
Jupiter moving on their unperturbed Keplerian orbits.

We also find three body resonances with arguments containing
\be %$
3\lambda_S-5\lambda_U-7\lambda_N+7g_6t+(2-q)s_7t+qs_8t,
\ee %$
at $19.2187\hbox{ AU}\lesssim a_U\lesssim19.2195\hbox{ AU}$.
The strength of the resonance is smaller than that of the resonance
involving Jupiter by the ratio
$(\mu_N/\mu_J)(\epsilon_{JS}/\epsilon_{SU})\approx 3\times10^{-3}$. 

{\it Overlapping resonances.} The overlap of the individual resonances
produces chaos among the outer planets. 
The width of a typical component resonance is
\be \label{width}%$
{\Delta a\over a_U}=8
\sqrt{(6-p)\phi^{(7,1)}_{6-p,p,0,0}\phi^{(2,5)}_{2,1,0,0}{\alpha\over3\epsilon_{JS}}
\mu_J\mu_Se_J^{5-p}e_U^pe_S^2}\approx2\times
10^{-6},
\ee %$
or  $\Delta a\approx8\times10^{-5}$ AU. We must substitute
powers of either $e_{55}$ or $e_{56}$ for $e_J^{5-p}$, depending on
the resonant argument. This resonance width 
is comparable to the radius of Uranus. The libration period is
\be %$
T_0=T_U\Bigg/\sqrt{147(6-p)\phi^{(7,1)}_{6-p,p,0,0}\phi^{(2,5)}_{2,1,0,0}
{\alpha\over\epsilon_{JS}}\mu_J\mu_Se_J^{5-p}e_U^pe_s^2}\approx 10^7
{\rm years}.
\ee %$

The precession frequencies $g_5$ and $g_7$ determine the distance between
the component resonances; we find 
\be \label{separation}%$
{\delta a\over a_U}\approx {4\pi\over 21}\left({g_5-g_7\over
n_U}\right)\approx 7\times10^{-6}
\ee %$

The stochasticity parameter is 
\be \label{stochastic}%$
K\equiv\left(\pi{\Delta a\over \delta a}\right)^2.
\ee %$
Using Eqns. (\ref{width}) and (\ref{separation}) in Eqn. (\ref{stochastic}),
we see that $K\gtrsim1$, so the motion is marginally chaotic.  Then
the Lyapunov time (the inverse of the Lyapunov exponent) is given by
$T_L\lesssim T_0$ \cite{MH}. 

The chaotic nature of the system ensures that the angles in the
perturbing potential (\ref{potential357}) experienced by Uranus are
essentially random variables. These chaotic perturbations force
Uranus's $e$ to undergo a random walk, exploring all values
between 0 and $e_{cross}\approx0.5$; for $e>e_{cross}$ Uranus will
suffer close encounters with Saturn, and may be ejected from the solar
system. The time for this to occur is of order \cite{MH}
\be %$
T_{cross}\approx 6\times10^{17}\left({0.05\over e_{cross}
}\right)^p{\rm years}
\ee %$
where $p$ is the exponent of $e_U$ in Eqn. (\ref{potential357}).  This
estimate is uncertain by a large factor, possibly by one or two orders
of magnitude, but it is clear that Uranus will be with us for a long
time.  The resonance closest to the actual value of $a_U$ has $p=0$.

The discovery that the great inequality was due to the $2:5$ near
resonance between Jupiter and Saturn clearly had a strong affect on
Laplace's views regarding determinism.  We find it ironic that the
$2:5$ resonance plays such a strong role in producing chaos among the
outer planets, thereby placing a limit on our ability to state the
positions of the Jovian planets in the distant future. The fact that
Laplace was the first astronomer to identify a three body resonance in
the solar system, involving three of the Galilean satellites, only
heightens the irony. More recently, three body resonances were shown
to be responsible for much of the chaos seen in integrations of
asteroids \cite{asteroids}.

{\bf Numerical integrations.}  In order to test our theory, we have
integrated the equations of motion for the four Jovian planets using a
symplectic integrator \cite{WH}. We chose this  simplified model
rather than including all nine planets in order to
isolate the effects of the giant planets. To account in a crude way
for the effects of the terrestrial planets, we enhanced the mass of
the sun by the their mass, roughly a part in $6\times10^{-6}$. This
ensures that the location of resonances between the Jovian planets is
shifted by an amount which is second order in this mass ratio, roughly
$3\times10^{-11}$. This is much smaller than the uncertainty in the
orbital elements of the planets. The orbital elements, which provide the
initial conditions for our integrations, are known to a relative
accuracy of a few parts in 10 million. For example, $\Delta a/a\sim2\times10^{-7}$ (600km for Uranus)
\cite{standish}, much smaller than the size of the resonances.

To determine whether the evolution was chaotic, we measured
the Lyapunov time by comparing pairs of integrations in which the
initial conditions differed by 1.5 millimeters in the $x$ coordinate
of Uranus. Using the DE200 ephemeris from JPL, we confirm the result
of Sussman and Wisdom \cite{SW} that the four Jovian planets are
chaotic. We find a Lyapunov time of about 7 million years, consistent
with our analytic result and with Sussman and Wisdom's result of about
5 million years, given that it is difficult to measure Lyapunov times
with an accuracy much better than a factor of 2.

To check the robustness of this conclusion, we have carried out
integrations in which we varied the initial $a_U$ in ten steps of
300km; the largest displacement was $\pm1,500$km, about twice the
uncertainty in the JPL ephemeris. We employed symplectic correctors
\cite{WHT} to
ensure that the relative energy errors were less than $10^{-9}$, much
smaller than the uncertainties in the initial conditions. In all these
integrations we found that the orbits were chaotic.

To test the prediction that the motion is marginally chaotic, we
carried out various surveys of the dynamics of the Jovian planets in
which all the initial orbital elements except $a_U$ were held fixed
\cite{quinlan}. The integration time in each survey was 200 million
years. In our first survey we varied the initial value of $a_U$ in
steps of 0.01AU between 18.9789 and 19.3990AU. We found that between
19.18 and 19.399 AU more than $80\%$ of the orbits are regular.
Subsequently we conducted a survey in which $a_U$ was varied in steps
of 0.0001AU between 19.2141 and 19.2209.  The resulting Lyapunov times
are plotted as a function of the initial semimajor axis $a_U$ in
Figure 1. We plot a point at $10^8$ years, corresponding to the
integration time, if the orbit appeared to be regular. The location of
our solar system as represented in the DE200 ephemeris is indicated by
the vertical line in the 
figure. From 18.9789 to 19.15 AU we find a strongly chaotic
region, with Lyapunov times ranging from 25,000 to 2 million
years. Examination of the resonant argument
$\lambda_U-2\lambda_N+\varpi_N$ reveals that from
18.9789 to about 19.13 AU our pseudo-Uranus is in a $1:2$ mean motion
resonance with Neptune. From 19.13 to 19.17 AU pseudo-Uranus is in the
$7:1$ mean motion resonance with Jupiter described by
Eqn. (\ref{seven2one}), with a Lyapunov time ranging upwards from
100,000 years. There are four other chaotic regions visible in  Figure 1,
centered at $a_U$ of 19.219, 19.26, 19.29, and 19.34 AU. All of these
regions are associated with three-body resonances.

The dynamics in the region from 19.21 to 19.225 AU (Figure 2) is
controlled by the $3\lambda_J-5\lambda_S-7\lambda_U$ three-body
resonance described in Eqn. (\ref{potential357}). We can see the
effects of the individual resonant terms. For $a_U<19.218\hbox{ AU}$
the resonances are isolated by regular regions, indicating that the
resonance widths are slightly smaller than the distance between
resonances. For $a_U\gtrsim19.218\hbox{ AU}$ nearly all the orbits
have finite Lyapunov times, indicating that the individual resonances
overlap completely. Figure 3 shows the resonant angle
$3\lambda_J-5\lambda_S-7\lambda_U+3g_5t+6g_6t$ (the $q=0$ case of
eq. (\ref{resonance})) for $a_U=19.21908$, about one planetary radius
larger than the value of $a_U$ used in the DE200 ephemeris. It
alternates between libration, with a period of about 20 million years,
and rotation, indicating that the orbit is crossing the separatrix of
the resonance and confirming the chaotic nature of the orbit. 
In addition to the $3\lambda_J-5\lambda_S-7\lambda_U$ resonance, there
is a resonant term involving Saturn, Uranus, and Neptune. Our
calculations suggest that this resonance is responsible for the
chaotic zones at $19.29$ and  $19.34$, and plays a strong role in the
chaotic zone at $19.26$.

{\it Integrations of simpler models.} In another survey, we set $i=0$
for all four Jovian planets and again 
varied $a_U$ in steps of 0.01 between 18.9789 and 19.3990 AU, and in
steps of 0.0001 between 19.2141 and 19.2209AU. The general appearance
is similar to that of Figure 1, showing that inclination resonances
are not essential to produce chaos among the Jovian planets. However,
the chaotic region near $a_U=19.219\hbox{ AU}$ is not quite so
extensive, and the resonances appear to be isolated, like those with
$a_U<19.218\hbox{ AU}$ in Figure 2. In yet another survey, we removed
Neptune. The chaotic region at $a_U\approx19.00\hbox{ AU}$ vanishes,
but the chaos associated with $7\lambda_U-\lambda_J$
remains. Similarly, the chaos at $a_U=19.29\hbox{ AU}$ and
$19.34\hbox{ AU}$ is no longer present. However a chaotic region at
$a_U=19.219\hbox{ AU}$ and a very small chaotic region at $19.25\hbox
{ AU}$ remain. The feature near $a_U=19.219\hbox{ AU}$ is even less
extensive than in the planar case, indicating that the effects of the
$3\lambda_S-5\lambda_U-7\lambda_N$ resonance are more important than
the effects of inclination resonances involving Jupiter. Finally, a
survey in which Neptune is removed and the remaining Jovian planets
orbit in the same plane reveals no chaotic motion outside the
$7\lambda_U-\lambda_J$ resonance. Apparently eccentricity resonances
involving only the inner three Jovians do not quite overlap. They must
act in concert either with inclination resonances or with three-body
resonances involving Neptune to produce detectable chaotic regions.

{\bf The epoc of resonance capture}.
Uranus probably did not form in the current resonance. Planet
formation is believed to occur in disks around young stars. Evidence
for such disks, which have lifetimes around ten million years, is now
abundant, including visible, infrared and millimeter observations of
disks around young stars \cite{Disks}. The observations show that the
disks contain both gas and particulate matter. The existence of our
own asteroid and Kuiper belts, as well as of comets, suggest that
protostellar disks contain larger bodies as well. Current
understanding of the planet formation process suggests that planets
migrate over substantial distances early in the history of a planetary
system. Goldreich and Tremaine \cite{GT80} showed that torques
produced by interactions between a gas disk and a planet can cause
large scale planet migrations on timescales of tens to hundreds of
thousands of years. Interactions between asteroids or comets and
planets can also cause planet migrations \cite{FI}. The recent
discovery \cite{extra_solar} of Jupiter-mass objects in short period
(4 day) orbits around nearby stars strongly suggest that planet
migration is common.

We can compare the torques exerted on Uranus by the different processes.
Jupiter and Saturn currently exert a resonant torque on Uranus given by
\be %$
T_{res}\approx 100\left({GM_\odot M_U\over a_U}\right)\mu_J\mu_S
\left({e_J^{5-p}e_U^pe_S^2\over \epsilon_{JS}}\right).
\ee %$
The torque exerted on proto-Uranus by the gas disk in which it formed
is
\be %$
T_{gas}\approx 5.6\left({GM_\odot M_U\over a_U}\right)\mu_U\mu_g m_{max}^3
\ee %$
\cite{GT80}.  In this expression the quantity $m_{max}$ is a measure
of the gap in the gas disk produced by Uranus. If no such gap formed,
the torque produced by the gas disk is even larger. The minimum mass
of the solar nebula is about 10 Jupiter masses, so $\mu_g\equiv M_{\rm
gas\ disk}/M_\odot\approx 0.01$. The torque produced by interactions
between Uranus and a planetesimal disk is
\be %$
T_{planetesimal}\approx\left({GM_\odot M_U\over a_U}\right)
\left({M_d\over M_U}\right)\left({T_U\over T_{clear}}\right),
\ee %$
where $M_d$ is the total mass of the planetesimals that interact with
Uranus, $T_U\approx 80$ years is the orbital period of Uranus, and
$T_{clear}\approx 10^7$ years is the time for Uranus to clear the
planetesimal disk. In units of $GM_\odot M_U/a_U$ the torques are
$T_{res}\sim 10^{-11}$, $T_{gas}\sim10^{-3}$, and
$T_{planetesimal}\sim10^{-6}M_d/M_U$. The planets remain 
in resonance only if $T_{res}\gtrsim T_{gas}$ and $T_{res}\gtrsim
T_{planetesimal}$. Clearly, Uranus must have been 
trapped in the resonance after the gas disk dissipated. Similarly,
most of the planetesimal disk must be removed before the final
trapping can occur.

% references
\vfil\eject
\begin{planotable}{lcccc}
%\tablewidth{470pt}
\tablecaption{Masses, in units of the solar mass $M_\odot$, and the
current semimajor axes $a$ (in AU), eccentricities $e$ and
inclinations $i$ of the orbits of the giant planets. Data taken from
JPL ephemeris DE200.}
\tablehead{
\colhead{Planet}		& \colhead{$\mu\equiv M/M_\odot$}	&
\colhead{$a$ (AU)  } & \colhead{$e$}   &
\colhead{$i$ (radians) }
}
\startdata
Jupiter	& $9.548\times10^{-4}$	& $5.207$	& $0.04749$ 	& $0.02277$ \cr
Saturn	& $2.859\times10^{-4}$	& $9.553$	& $0.05274$	& $0.04338$	\cr
Uranus	& $4.355\times10^{-5}$	& $19.219$	& $0.04641$	& $0.01348$	\cr
Neptune	& $5.178\times10^{-5}$	& $30.111$	& $0.00820$	& $0.03089$	\cr
\end{planotable}
\begin{planotable}{lcccc}
%\tablewidth{470pt}
\tablecaption{Orbital frequencies of the giant planets, in cycles per
day. Data from our numerical integrations.} 
\tablehead{
\colhead{Planet/mode}		&
\colhead{$n/2\pi$ (days$^{-1}$)  } & \colhead{$g$ (days$^{-1}$) }   &
\colhead{$s$ (days$^{-1}$) }
}
\startdata
5 \phantom {Jupiter}& $2.308\times10^{-4}$	& $8.967\times10^{-9}$ 	& $0.0$ \cr
6	& $9.294\times10^{-5}$	& $5.965\times10^{-8}$	& $-5.564\times10^{-8}$	\cr
7	& $3.259\times10^{-5}$	& $6.520\times10^{-9}$	& $-6.328\times10^{-9}$	\cr
8	& $1.662\times10^{-5}$	& $1.420\times10^{-9}$	& $-1.460\times10^{-9}$	\cr
\end{planotable}
\vfil\eject

%Figure captions
\vfil\eject

Fig. 1. The Lyapunov time $T_L$ as a function of initial $a_U$. The initial orbital
elements of the planets are taken from DE200, except for $a_U$, which
is varied. There are chaotic two body resonances at $a_U\approx 19.00$ and
$19.12$AU involving Neptune and Jupiter, respectively. There are also
chaotic regions associated with three body mean motion resonances at
$a_U\approx19.18$, $19.25$, $19.29$, and $19.34$ AU. These involve
either Jupiter, Saturn, and Uranus, or Saturn, Uranus and Neptune. The
solid vertical line shows the actual location of Uranus.

Fig. 2. A close up of Figure 1 around the 
actual value of $a_U$. Between $19.216$ and $19.218$ AU we find the
individual eccentricity resonances associated with the resonant
argument $3\lambda_J-5\lambda_S-7\lambda_U+qg_5+7g_6t+(2-q)g_7t$,
which do not quite overlap. The resonances associated with the
argument $3\lambda_J-5\lambda_S-7\lambda_U+qg_5+6g_6t+(3-q)g_7t$ lie
between $19.218$ and $19.221$ AU. 

Fig. 3. The resonant argument
$3\lambda_J-5\lambda_S-7\lambda_U+3g_5t+6g_6t$ in the case
$a_U=19.21908$, about one planetary radius larger than the actual
value of $a_U$. The libration period is $T_0\approx 20,000,000$ 
years. A transition from libration to rotation occurs near 60 million
years. A longer lasting transition from libration to rotation occurs
at 160 million years. The Lyapunov time was measured to be about 7
million years.

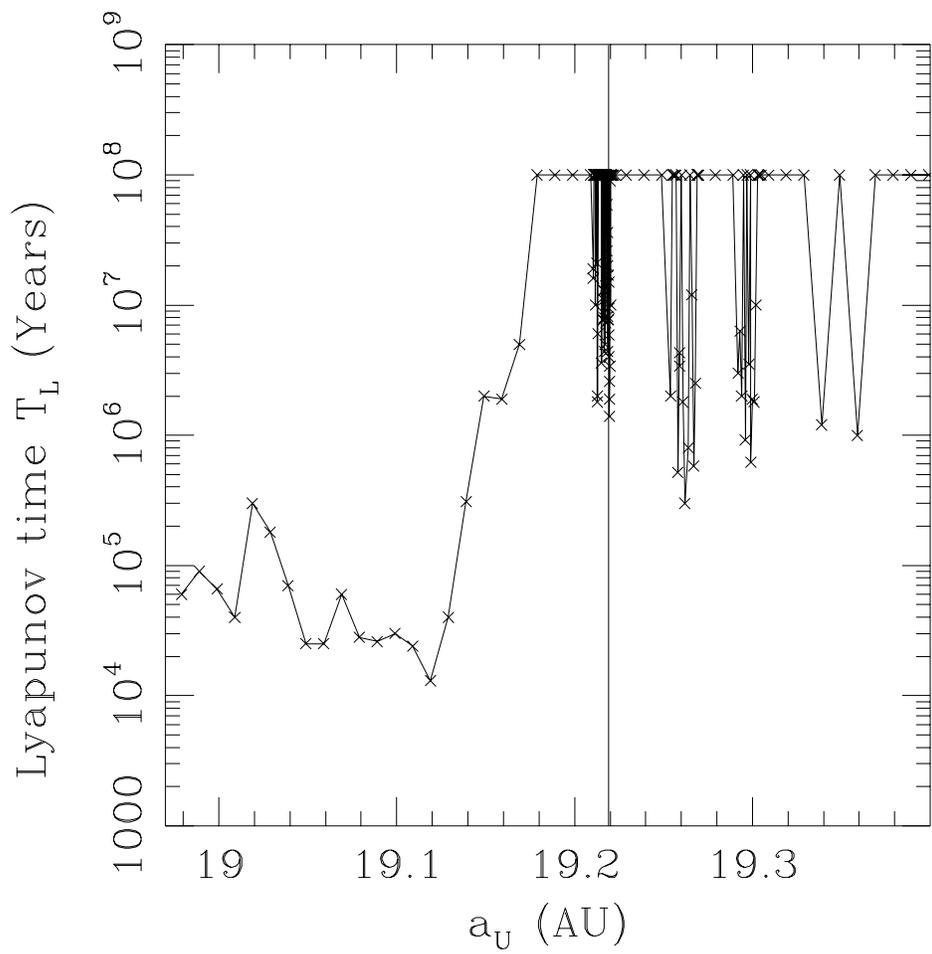
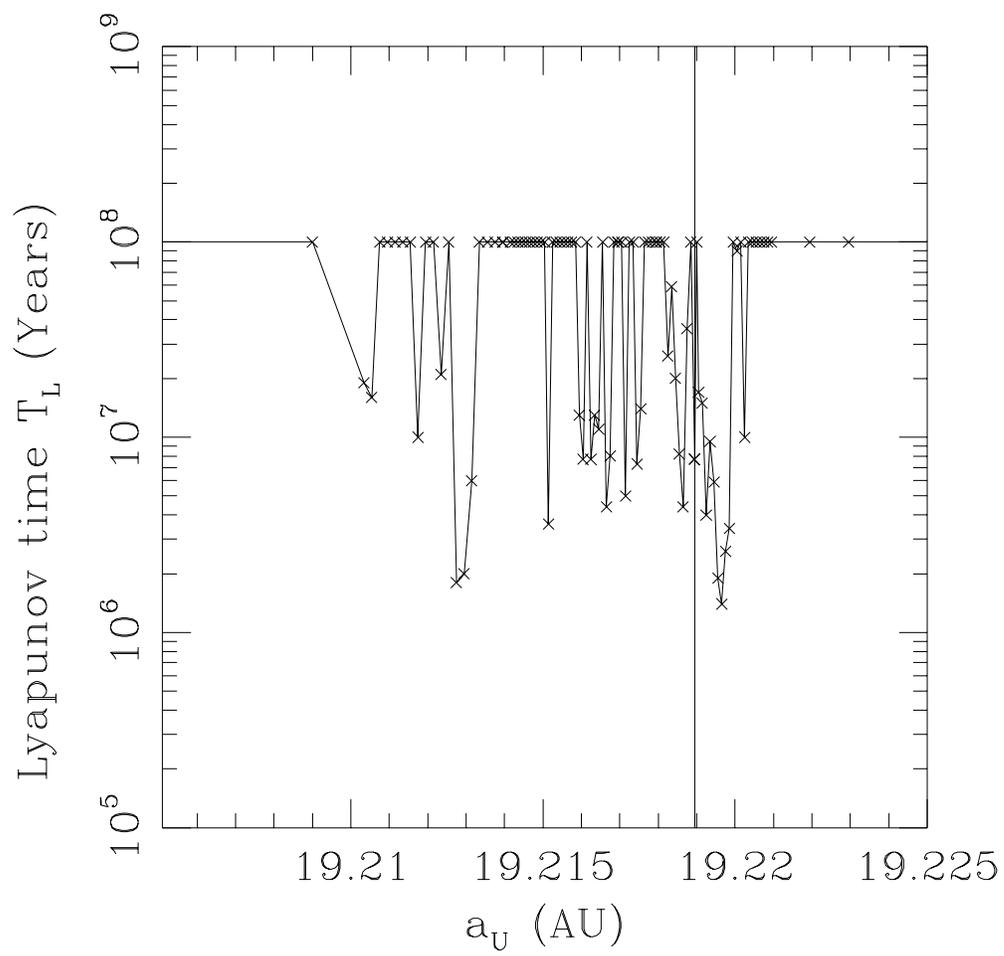
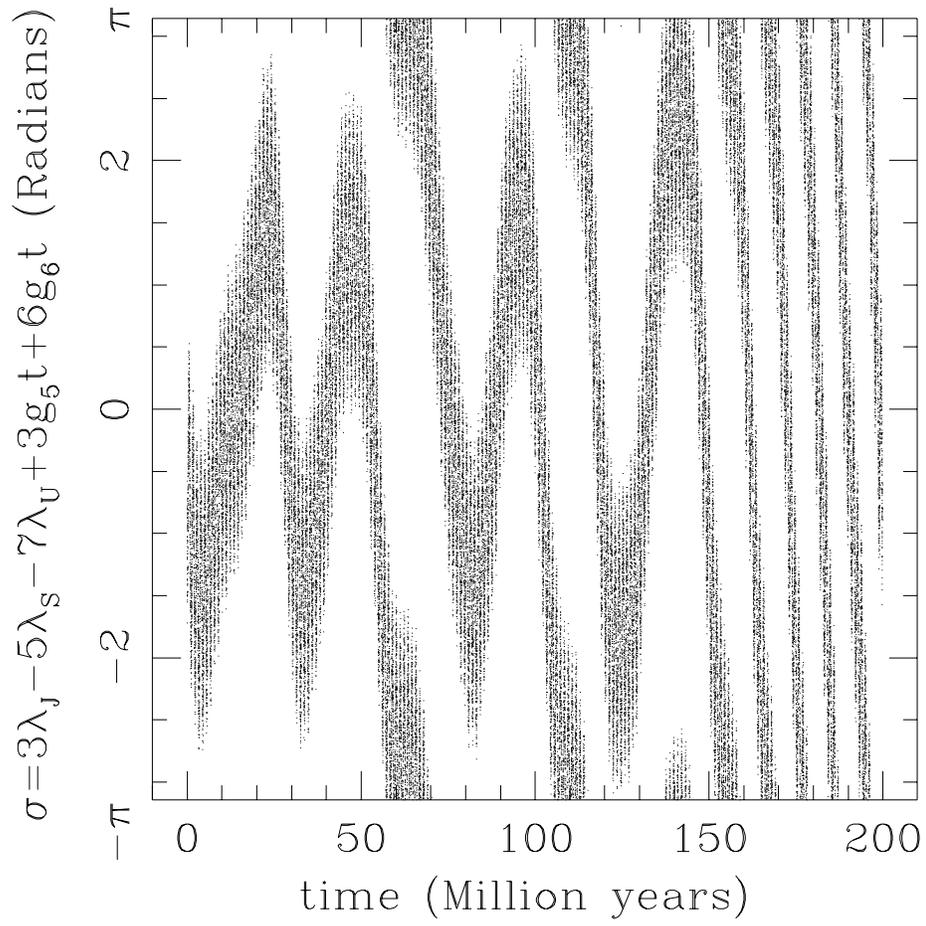
\end{document}